\documentclass[letterpaper, twocolumn, pra]{revtex4}
\usepackage{graphicx,amsmath, amssymb, bm, epstopdf,dsfont, enumerate}
\DeclareMathOperator{\sinc}{sinc}
\begin{document}

\title{Angular Schmidt spectrum of entangled photons: derivation of an exact formula and experimental characterization for non-collinear phase matching}

\author{Girish Kulkarni, Lavanya Taneja, Shaurya Aarav, and Anand K. Jha}

\email{akjha9@gmail.com}

\affiliation{Department of Physics, Indian Institute of Technology Kanpur, Kanpur, UP 208016, India}

\date{\today}

\begin{abstract}

We derive an exact computationally-efficient formula for the angular Schmidt spectrum of  orbital angular momentum (OAM)-entangled states produced by parametric down-conversion (PDC). Our formula yields the true spectrum and does not suffer from convergence issues arising due to infinite summations, as has been the case with previously derived formulas. We use this formula to experimentally characterize the angular Schmidt spectrum of entangled photons produced by PDC with non-collinear phase matching. We report measurements of very broad angular Schmidt spectra, corresponding to the angular Schmidt numbers up to 229. Our studies can have important implications for OAM-based quantum information applications.

\end{abstract}

\maketitle

\section{Introduction}

High-dimensional quantum information protocols offer
many distinct advantages in terms of security
\cite{karimipour2002pra,cerf2002prl,nikolopoulos2006pra},
supersensitive measurements \cite{jha2011pra2}, violation of bipartite Bell's inquality \cite{collins2002prl, leach2009optexp,vetesi2010prl}, enhancement of
entanglement via concentration \cite{vaziri2003prl}, and implementation of quantum coin-tossing protocol \cite{terriza2005prl}. After it was shown that the orbital angular momentum (OAM) of a photon provides a high-dimensional basis \cite{allen1992pra, barnett1990pra, yao2006opex}, the OAM-entangled states of signal and idler photons produced by parametric down-conversion (PDC) have become a natural choice for high-dimensional quantum information applications. To this end, there have been intense research efforts, both theoretically \cite{law2004prl,
torres2003pra, miatto2011pra, yao2011njp, jha2011pra, miatto2012epjd, zhang2014pra} and experimentally \cite{mair2001nature, pors2008prl, jha2010prl,
peeters2007pra, pires2010prl, giovannini2012njp, kulkarni2017natcomm}, for the precise characterization of high-dimensional OAM-entangled states produced by PDC. Although a general OAM-entangled state requires the full state tomography for its characterization, the experimentally relevant case of OAM-entangled states produced using a Gaussian pump beam can be characterized by measuring just the angular Schmidt spectrum  \cite{law2004prl,
torres2003pra, pires2010prl}, which is defined as the probability $S_l$ of signal and idler photons getting detected with OAMs $l\hbar$ and $-l\hbar$, respectively. 

The characterization of the angular Schmidt spectrum has been a very challenging problem. On the experimental front, several techniques have been developed for measuring the angular Schmidt spectrum. The first set of techniques is based on using fiber-based projective measurements \cite{mair2001nature, jha2010prl, pors2008prl, peeters2007pra, giovannini2012njp}. However, these techniques are very inefficient because the required number
of measurements scales with the size of the input spectrum. Furthermore, these techniques measure only the projected spectrum instead of the true spectrum \cite{qassim2014josab}. The second set of techniques is based on inferring the spectrum by measuring the angular coherence function \cite{pires2010prl, jha2011pra}. Although these techniques do measure the true spectrum, they either require a series of coincidence measurements and have strict interferometric stability requirements \cite{pires2010prl} or suffer from too much loss \cite{jha2011pra}. More recently, an interferometric technique has been demonstrated that can measure the true angular Schmidt spectrum in a very efficient single-shot manner \cite{kulkarni2017natcomm}. On the theoretical front, Torres {\it et al.} have derived a formula for calculating the spectrum for collinear phase matching \cite{torres2003pra}. However, this formula involves a four-dimensional integration followed by two infinite summations over the radial indices. Although the summations have been shown to converge for certain set of experimental parameters, the convergence is not explicitly proved for an arbitrary set of parameters. Moreover, it is extremely inefficient to first calculate the contributions due to sufficiently large number of radial modes and then sum them over. Subsequent studies have analytically performed the four-dimensional integration for certain collinear phase-matching conditions \cite{miatto2011pra, yao2011njp}, but they still suffer from the same set of issues due to infinite summations. There has been a recent investigatoin by Zhang and Roux for the non-collinear phase matching condition \cite{zhang2014pra}, however, the angular Schmidt spectrum calculated in this study is only for a given pair of radial modes of the signal and idler photons, and therefore is not applicable to a generic experimental situation. 

Thus, although the past efforts have been able to greatly overcome the experimental challenges in measuring the true Schmidt spectrum, the theoretical challenge of deriving an exact formula has so far remained unresolved. In this article, we derive an exact formula for calculating the true angular Schmidt spectrum that does not suffer from the above mentioned issues since the infinite summations over radial modes are performed analytically. Moreover, our formula is valid for both collinear and non-collinear phase matching conditions. Using this formula, we report experimental characterizations of the angular Schmidt spectrum with various non-collinear phase-matching conditions. 

\section{Theory}

\subsection{Derivation of the general formula}

The state $|\psi_2\rangle$ of the down-converted photons is written in the transverse-momentum basis as \cite{hong1985pra}: 
\begin{align}
|\psi_2\rangle=\iint_{-\infty}^{\infty}  \Phi({\bm {q_s}, {\bm q_i}})|{\bm q_s}\rangle_s |{\bm q_i}\rangle_i d\bm q_s d\bm q_i, \label{state_momentum}
\end{align}
where, $s$, and $i$ stand for signal, and idler, respectively, and where $|{\bm q_{s}}\rangle$ and $|{\bm q_{i}}\rangle$ denote the states of the signal and idler photons with transverse momenta $\bm q_{s}$ and $\bm q_{i}$, respectively. $\Phi({\bm {q_s}, {\bm q_i}})$ is the wavefunction of the down-converted photons in the transverse-momentum basis; it depends on the detailed properties of the pump field, the nonlinear crystal, and the phase matching condition \cite{hong1985pra, torres2003pra, walborn2010physrep}. The state $|\psi_2\rangle$ can also be represented in the Laguerre-Gaussian (LG) basis \cite{torres2003pra, miatto2011pra, yao2011njp, jha2011pra} as:
\begin{align}
|\psi_2\rangle=\sum_{l_{s}}\sum_{l_{i}}\sum_{p_s}\sum_{p_i} C^{l_{s},p_{s}}_{l_i, p_i} |l_{s}, p_s\rangle_s|l_{i}, p_i\rangle_i. \label{state_LG}
\end{align}
Here $|l_{s}, p_s\rangle_s$ represents the state of the signal photon in the Laguerre-Gaussian (LG) basis defined by the OAM-mode index $l_{s}$ and the radial index $p_s$, etc. Using Eqs.~(\ref{state_momentum}) and (\ref{state_LG}), the complex coefficients $C^{l_s,p_s}_{l_i, p_i}$ can be written as,
\begin{align}
\!\!C^{l_{s},p_{s}}_{l_i,p_i}=\!\! \iint \!\! \Phi({\bm q_s}, {\bm q_i})LG^{*l_{s}}_{p_s}({\bm q_s})LG^{*l_{i}}_{p_i}({\bm q_i})d{\bm q_s} d{\bm q_i}. \label{schmidt-amp}
\end{align}
Here $LG^{l_s}_{p_s}({\bm q_s})=\langle {\bm q_s} | l_s, p_s \rangle$ is the momentum-basis representation of state $| l_s, p_s \rangle_s$ \cite{torres2003pra, miatto2011pra}.  Transforming to the cylindrical coordinates, we write $C^{l_s,p_s}_{l_i,p_i}$ as,
\begin{multline}
C^{l_{s},p_{s}}_{l_i,p_i}=\iint_{0}^{\infty}\iint_{-\pi}^{\pi}  \Phi(\rho_s, \rho_i, \phi_s,\phi_i) \\ \times LG^{*l_s}_{p_s}(\rho_s, \phi_s) LG^{*l_i}_{p_i}(\rho_i, \phi_i) \rho_s\rho_i d\rho_s d\rho_i d\phi_s d\phi_i, \label{schmidt-amp-LG}
\end{multline}
where ${\bm q_s}\equiv(q_{sx}, q_{sy})=(\rho_s\cos\phi_s,  \rho_s\sin\phi_s)$, ${\bm q_i}\equiv(q_{ix}, q_{iy})=(\rho_i\cos\phi_i,  \rho_i\sin\phi_i)$, $d{\bm q_s}=\rho_s d\rho_s d\phi_s$, and $d{\bm q_i}=\rho_i d\rho_i d\phi_i$. The probability $P^{l_{s}}_{l_{i}}$, that the signal and idler photons are detected with OAMs $l_s\hbar$ and $l_i\hbar$, respectively, is calculated by summing over radial indices:
\begin{align}
P^{l_{s}}_{l_{i}}=\sum_{p_s=0}^{\infty}\sum_{p_i=0}^{\infty}|C^{l_{s},p_{s}}_{l_i,p_i}|^2. \label{OAM_prob}
\end{align}
Eqs.~(\ref{schmidt-amp-LG}) and (\ref{OAM_prob}) were used in Refs.~\cite{torres2003pra, miatto2011pra, yao2011njp} for calculating the specta of OAM-entangled states. We note that in order to calculate the angular Schmidt spectrum using the above formula one needs to first choose a beam waist for the signal and idler LG bases in Eqs.~(\ref{schmidt-amp-LG})and then perform the summations in Eq.~(\ref{OAM_prob}) over a sufficiently large number of modes. As a result, even for certain collinear phase-matching conditions, in which the four-dimensional integral can be analytically performed \cite{miatto2011pra, yao2011njp}, the above formula suffers from convergence issues. 

We next present the derivation of a formula for the angular Schmidt spectrum that neither requires a beam waist to be chosen nor involves infinite summations and is  applicable to both collinear and non-collinear phase matching conditions. To this end, we first rewrite Eq.~(\ref{OAM_prob}) using the relation $LG^{l_s}_{p_s}(\rho_s, \phi_s)=LG_{p_s}^{l_s}(\rho_s)e^{il_s\phi_s}$, etc., as
\begin{align}
P^{l_s}_{l_i}
&=\iiiint_{0}^{\infty}\iiiint_{-\pi}^{\pi}  \Phi(\rho_s, \rho_i, \phi_s,\phi_i) \Phi^*(\rho'_s, \rho'_i, \phi'_s,\phi'_i)
\notag\\ 
& \times\sum_{p_s=0}^{\infty} LG^{*l_s}_{p_s}(\rho_s) LG^{l_s}_{p_s}(\rho'_s) 
\sum_{p_i=0}^{\infty} LG^{*l_i}_{p_i}(\rho_i) LG^{l_i}_{p_i}(\rho'_i) \notag \\ &\times e^{+i(l_s\phi_s+l_i\phi_i)} e^{-i(l_s\phi'_s+l_i\phi'_i)} \notag \\
& \times \rho_s\rho_i \rho'_s\rho'_i
d\rho_s d\rho_i d\rho'_s d\rho'_i d\phi_s d\phi_i d\phi'_s d\phi'_i. 
\end{align}
We then use the identity $\sum_{p=0}^{\infty} LG^{p}_{l}(\rho)LG^{*^{p}}_{l}(\rho')=(1/\pi)\delta(\rho^2-\rho'^2)$ over indices $p_{s}$ and $p_{i}$ and obtain
\begin{align}
P^{l_s}_{l_i} &=\iiiint_{0}^{\infty}\iiiint_{-\pi}^{\pi}  \Phi(\rho_s, \rho_i, \phi_s,\phi_i) \Phi^*(\rho'_s, \rho'_i, \phi'_s,\phi'_i) \notag\\ 
&\times\frac{1}{\pi^2}\delta(\rho_s^2-\rho'^2_s)\delta(\rho_i^2-\rho'^2_i) e^{+i(l_s\phi_s+l_i\phi_i)} e^{-i(l_s\phi'_s+l_i\phi'_i)} \notag\\
&\times\rho_s\rho_i \rho'_s\rho'_i d\rho_s d\rho_i d\rho'_s d\rho'_i d\phi_s d\phi_i d\phi'_s d\phi'_i .
\end{align}
After evaluating the delta function integrals and rearranging the remaining terms, we obtain
\begin{align}
P^{l_s}_{l_i} =\frac{1}{4\pi^2}\iint_{0}^{\infty} &{\Big |}\iint_{-\pi}^{\pi}  \Phi(\rho_s, \rho_i, \phi_s,\phi_i)  \notag \\
& \times e^{i(l_s\phi_s+l_i\phi_i)}d\phi_s d\phi_i  {\Big |}^2 
\rho_s\rho_i d\rho_s d\rho_i. \label{angular_schmidt_spectrum} 
\end{align}
Now, we take up the most common experimental situation in which the OAM remains conserved during down-conversion, that is, $l_p=l_s+l_i$, which for a Gaussian pump beam with $l_p=0$ implies that $l_s=-l_i=l$  \cite{mair2001nature}. In these situations, the down-converted two-photon state $|\psi_2\rangle$ of Eq.~(\ref{state_LG}) takes the following form \cite{law2004prl, torres2003pra, miatto2011pra, yao2011njp, jha2011pra}: $|\psi_2\rangle=\sum_{l}\sum_{p_s}\sum_{p_i}C^{l,p_{s}}_{-l,p_i} |l, p_s\rangle_s|-l, p_i\rangle_i$, which, when written with only the OAM-mode index as the label for the state, takes the Schmidt decomposed form:
$|\psi_2\rangle=\sum_{l=-\infty}^{\infty}\sqrt{S_l}|l\rangle_s|-l\rangle_i$. The corresponding angular Schmidt spectrum $S_l= P_l^{-l}$ is the probability that the signal and idler photons have OAMs $l\hbar$ and $-l\hbar$, respectively, and using Eq.~(\ref{angular_schmidt_spectrum}) it can be written as 
\begin{align}
S_{l} = \frac{1}{4\pi^2}\iint_{0}^{\infty} &{\Big |}\iint_{-\pi}^{\pi}  \Phi(\rho_s, \rho_i, \phi_s,\phi_i)  \notag \\
& \times e^{il(\phi_s-\phi_i)}d\phi_s d\phi_i  {\Big |}^2  \rho_s\rho_i d\rho_s d\rho_i . \label{angular_schmidt_spectrum2} 
\end{align} 
Equations (\ref{angular_schmidt_spectrum}) and (\ref{angular_schmidt_spectrum2}) are the main theoretical results of this article. While Eq.~(\ref{angular_schmidt_spectrum}) provides a formula for calculating the probability $P^{l_s}_{l_i}$ that the signal and idler photons are detected with OAMs $l_s\hbar$ and $l_i\hbar$, respectively, Eq.~(\ref{angular_schmidt_spectrum2}) calculates the angular Schmidt spectrum. In contrast to the previously obtained formulas \cite{torres2003pra, miatto2011pra, yao2011njp, zhang2014pra}, Eqs.~(\ref{angular_schmidt_spectrum}) and (\ref{angular_schmidt_spectrum2}) neither require a beam waist to be chosen nor involve infinite summations. As a result, these formulas can provide improvement of several orders of magnitude in the spectrum computation time. Moreover, unlike the non-collinear phase-matching results in Ref.~\cite{zhang2014pra}, which is applicable only for a given pair of radial modes of the signal and idler photons, these formulas are applicable to a generic set of non-collinear phase matching conditions and geometries. We note that although the above formulas do not have any convergence issue arising due to infinite summations, the definite integrals might have convergence issues for some arbitrary functional form of $\Phi(\rho_s, \rho_i, \phi_s,\phi_i)$. However, we do not expect such convergence issues for the commonly encountered forms of $\Phi(\rho_s, \rho_i, \phi_s,\phi_i)$ for collinear and non-collinear phase matching conditions. In order to illustrate this and to describe our experiments presented later, we next derive the momentum-space wavefunction $\Phi(\rho_s, \rho_i, \phi_s,\phi_i)$ for the case of collinear type-I down-conversion and calculate the angular Schmidt spectrum. 

\begin{figure}[t!]
\includegraphics{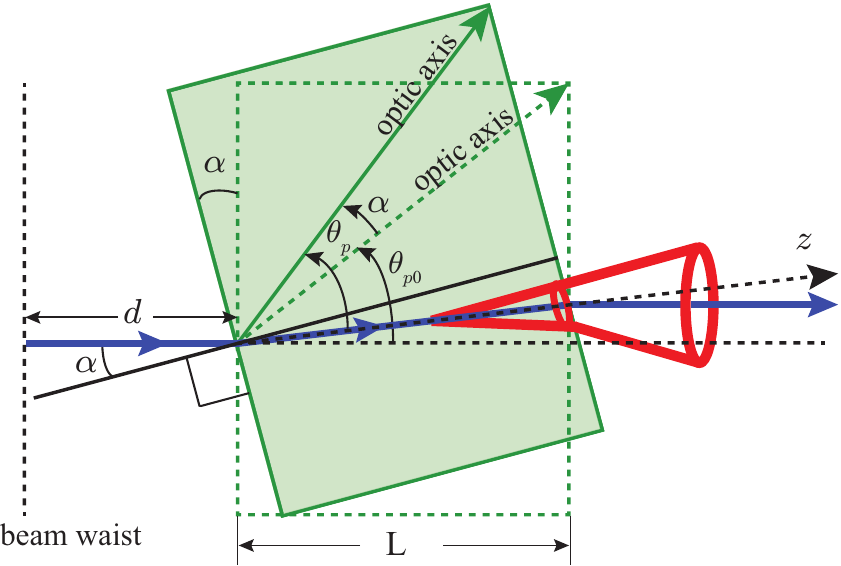}
\caption{(color online) Schematic of phase matching in PDC.}
\label{fig1}
\end{figure}

\subsection{The special case of a Gaussian pump beam}

Let us consider the situation shown in Fig.~\ref{fig1}. A Gaussian pump beam undergoes Type-I PDC inside a nonlinear crystal of thickness $L$. We take the pump photon to be extra-ordinary polarized and the signal and idler photons to be ordinary polarized. The beam waist of the pump field is located at a distance $d$ behind the front surface of the crystal. The crystal is rotated by an angle $\alpha$ with respect to the incident direction of the pump beam, and the $z$-axis is defined to be the direction of the refracted pump beam inside the crystal. The angles that the optic axes of the unrotated and rotated crystals make with the pump beam inside the crystal are denoted by $\theta_{p0}$ and $\theta_p$, respectively. Using Fig.~\ref{fig1}, one can show that  
\begin{align}
\theta_p=\theta_{p0}+\sin^{-1} \left(\sin\alpha /\eta_p\right), \label{theta_p}
\end{align}
where $\eta_p$ is the refractive index of the extraordinary pump photons. By changing $\theta_p$, one can go from collinear to non-collinear down-conversion.  The wavefunction $\Phi({\bm {q_s}, {\bm q_i}})$ of the down-converted photons in the transverse-momentum basis at the exit surface inside the crystal is written as \cite{hong1985pra, torres2003pra, walborn2010physrep}: 
\begin{figure}[t!]
\includegraphics{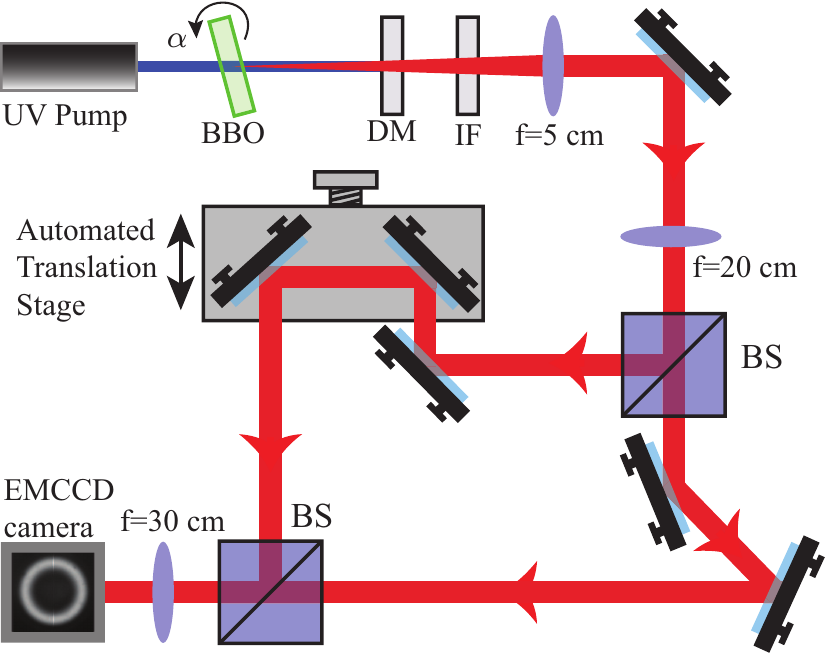}
\caption{(color online) Experimental setup for measuring the angular Schmidt spectrum. BBO: $\beta$-Barium Borate crystal; DM: Dichroic mirror; IF: 10-nm wavelength-bandwidth interference filter; BS: beam splitter.}
\label{fig2}
\end{figure}
\begin{align}
\Phi({\bm {q_s}, {\bm q_i}})=&A V({\bm q_s}+{\bm q_i})e^{ik_{pz} d} \notag \notag\\
&\times\sinc\left(\frac{\Delta k_z L}{2}\right)\exp\left(i\frac{\Delta k_z L}{2}\right),\label{wavefun_momentum}
\end{align}
Here, again, $p$, $s$, and $i$ stand for pump, signal, and idler, respectively; $A$ is a constant and $\mathrm{sinc}(x)\equiv\sin x/x$. We have used ${\bm k_j}\equiv (k_{jx}, k_{jy}, k_{jz})\equiv (q_{jx}, q_{jy}, k_{jz})\equiv ({\bm q_j}, k_{jz})$, with $j=p, s, i$, and $\Delta k_z=k_{pz}-k_{sz}-k_{iz}$. The quasi-monochromaticity condition is assumed for each of the signal, idler and pump photons with their central wavelengths given by $\lambda_s$, $\lambda_i$, and $\lambda_p$, respectively. In addition, the transverse size of the crystal is taken to be much larger compared to the spot-size of the pump beam, ensuring ${\bm q_p}={\bm q_s}+{\bm q_i}$. The quantity $V({\bm q_s}+{\bm q_i})e^{ik_{pz} d}$ is the spectral amplitude of the pump field at $z=0$, wherein
\begin{figure*}[t!]
\includegraphics{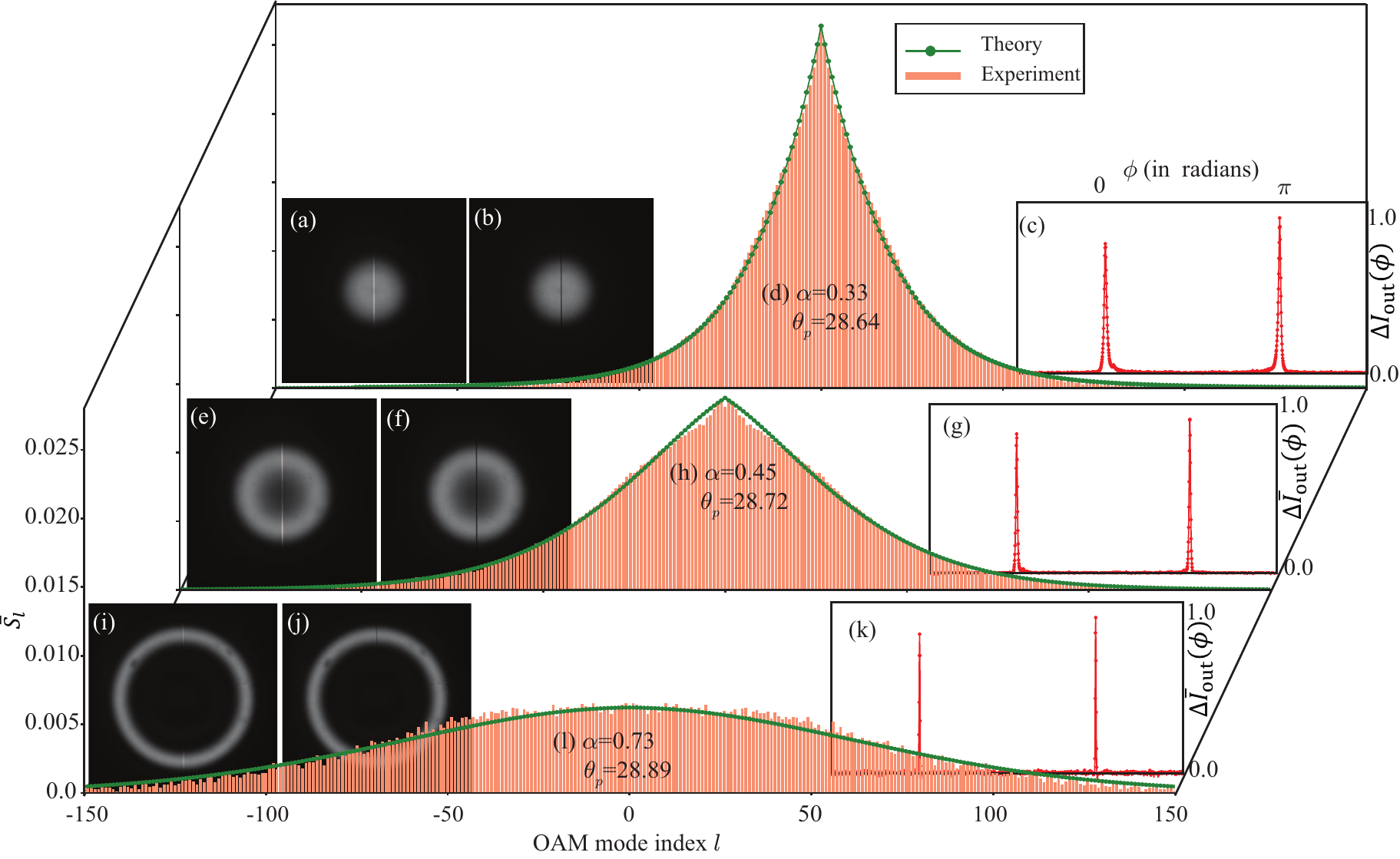}
\caption{(color online) (a), (b) The measured output interferograms at $\delta=\delta_c$ and $\delta=\delta_d$, respectively, (c) the difference $\Delta{\bar I}(\phi)$ in the azimuthal intensities of the two inteferograms, (d) The normalized measured spectrum as computed using Eq.~(\ref{measured_OAM_spectrum}) and the normalized theoretical spectrum as calculated using Eq.~(\ref{angular_schmidt_spectrum}), for $\alpha=0.33$ and $\theta_p=$28.64. (e), (f), (g),  (h) are the corresponding plots for $\alpha=0.45$ and $\theta_p=$28.72. (i), (j), (k), (l) are the corresponding plots for $\alpha=0.73$ and $\theta_p=$28.89.} \label{fig3} 
\end{figure*}
\begin{align}
V({\bm q_s}+{\bm q_i})= \frac{w_p}{\sqrt{2\pi}}\exp\left(-\frac{|{\bm q_s}+{\bm q_i}|^2 w_p^2}{4}\right)
\end{align}
is the spectral amplitude of the pump field at $z=0$ with $w_{p}$ being the width of the pump beam waist. We take the expressions for $k_{jz}$ from Ref.~\cite{walborn2010physrep} (a sign typo in the expression for $k_{pz}$ in Ref.~\cite{walborn2010physrep} has been corrected here):
\begin{align}
&k_{s_{z}}=\sqrt{ \left(2\pi n_{so}/\lambda_s\right)^2 -|{\bm q_s}|^2}, \notag \\  
&k_{i_{z}}=\sqrt{ \left(2\pi n_{io}/\lambda_i\right)^2 -|{\bm q_i}|^2},\ \ {\rm and} \ \notag \\
 &k_{p_{z}}= - \alpha_{p} q_{px}+\sqrt{ \left(2\pi \eta_{p}/\lambda_p\right)^2 -\beta_p^2 q_{px}^2-\gamma_p^2 q_{py}^2}, \label{kz}
\end{align}
where
\begin{align}
&\eta_{p}=n_{pe}\gamma_p ,  \notag \\
&\gamma_{p}=n_{po}/\sqrt{n_{po}^2\sin^2\theta_{p}+n_{pe}^2\cos^2\theta_{p}},  \notag\\
&\alpha_p =\frac{(n_{po}^2-n_{pe}^2)\sin\theta_{p}\cos\theta_{p}}{(n_{po}^2\sin^2\theta_{p}+n_{pe}^2\cos^2\theta_{p})},\notag\\
{\rm and} \ \ &\beta_{p}=\frac{n_{po}n_{pe}}{(n_{po}^2\sin^2\theta_{p}+n_{pe}^2\cos^2\theta_{p})}.\label{RI}
\end{align}
Here $n_{so}$ denotes the ordinary refractive index of the signal photon at wavelength $\lambda_{s}$, etc. The angular Schmidt spectrum $S_l$ can be evaluated by substituting into Eq.~(\ref{angular_schmidt_spectrum2}) from Eqs.~(\ref{wavefun_momentum}) through (\ref{RI}). We note that the formula in Eq.~(\ref{angular_schmidt_spectrum2}) represents angular Schmidt spectrum just inside the nonlinear crystal. Nevertheless, in situations in which $\alpha$ is of the order of only a few degrees, the angular Schmidt spectrum inside and outside the crystal can be taken to be the same.

Next, we use the experimental technique of Ref.~\cite{kulkarni2017natcomm} to characterize the angular Schmidt spectrum for non-collinear phase matching conditions and compare our experimental results with the theoretical predictions of Eq.~(\ref{angular_schmidt_spectrum2}). Figure \ref{fig2} shows our experimental setup. Following Ref.~\cite{kulkarni2017natcomm}, we first define the measured OAM spectrum as 
\begin{align}
\bar{S_l}\equiv\int_{-\pi}^{\pi}\Delta \bar{I}_{\rm out}(\phi)
e^{i2l\phi} d\phi, \label{measured_OAM_spectrum}
\end{align}
where $\Delta \bar{I}_{\rm
out}(\phi)=\bar{I}_{\rm out}^{ \delta_c}(\phi)-\bar{I}_{\rm out}^{
\delta_d}(\phi)$ is the difference in the azimuthal intensities $\bar{I}_{\rm out}^{ \delta_c}(\phi)$ and $\bar{I}_{\rm out}^{
\delta_d}(\phi)$ of the two output interferograms recorded  at $\delta=\delta_c$ and $\delta=\delta_d$, respectively, and where $\delta$ denotes the overall phase difference between the two arms of the interferometer \cite{kulkarni2017natcomm}. In situations in which the noises in the two interferograms are the same, it has been shown that $\bar{S_l}\propto S_l$, which implies that the normalized measured OAM-spectrum $\bar{S_l}$ is same as the true normalized OAM-spectrum $S_l$ \cite{jha2011pra, kulkarni2017natcomm}.

\section{Experimental Observations}

In the setup of Fig.~\ref{fig2}, an ultraviolet continuous-beam pump laser ($100$ mW) of wavelength $\lambda_{p}=405$ nm and beam-waist width $w_{p}=388$ $\mu$m was used to produce Type-I PDC inside a $\beta$-barium borate (BBO) crystal.  The beam waist of the pump field was located at $d=100$ cm behind the front surface of the crystal. The crystal was mounted on a goniometer which was rotated in steps of $0.04$ degrees in order to change $\alpha$ and thereby $\theta_p$. For a given setting of crystal and pump parameters, output interferograms and thereby the azimuthal intensities  $\bar{I}_{\rm out}^{\delta_c}(\phi)$ and $\bar{I}_{\rm out}^{\delta_d}(\phi)$ were recorded for two values of $\delta$, namely $\delta_{c}$ and $\delta_{d}$, which differed by about half a wavelength  \cite{kulkarni2017natcomm}. The recording of the interferograms was done using an Andor Ixon Ultra EMCCD camera ($512\times512$ pixels) with an acquisition time of $16$ seconds.  From a given pair of $\bar{I}_{\rm out}^{\delta_c}(\phi)$ and $\bar{I}_{\rm out}^{\delta_d}(\phi)$, $\Delta \bar{I}_{\rm out}(\phi)$ was obtained and the angular Schmidt spectrum was then estimated using Eq.~(\ref{measured_OAM_spectrum}). In our experiments, $\lambda_s=\lambda_i=810$ nm, $\lambda_p=405$ nm, and  $L=2$ mm. We have used the following refractive index values taken from Ref. \cite{eimerl1987jap}: $n_{po}=1.6923, n_{pe}=1.5680$ and $n_{so}=n_{io}=1.6611$.

\begin{figure}[t!]
\includegraphics{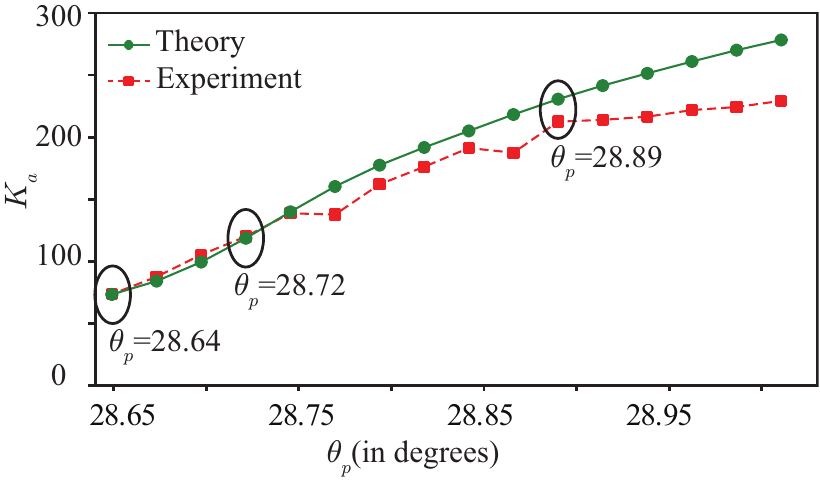}
\caption{(color online) Experimentally measured and theoretically estimated angular Schmidt number $K_a$ versus $\theta_p$.}
\label{fig4}
\end{figure}

Figure \ref{fig3} shows the details of our measurements for three different values of $\theta_p$. For each $\theta_p$, we have plotted the measured output interferograms at $\delta=\delta_c$ and $\delta=\delta_d$, the difference azimuthal intensity $\Delta{\bar I}_{\rm out}(\phi)$ along with the normalized spectrum as computed using Eq.~(\ref{measured_OAM_spectrum}) and the normalized theoretical spectrum as calculated using Eq.~(\ref{angular_schmidt_spectrum2}). The angular Schmidt number was calculated using the formula $K_a=1/\left(\sum_l {\bar S^2}_l\right)$. The experimentally measured angular Schmidt numbers along with the theoretical predictions at various $\theta_p$ values have been plotted in Fig.~\ref{fig4}. We note that for our theoretical plots, $\theta_{p0}$ was the only fitting parameter, and once it was chosen, the subsequent $\theta_p$ values were calculated simply by substituting the rotation angle $\alpha$ in Eq.~(\ref{theta_p}). We find that the angular Schmidt spectrum becomes broader with increasing non-collinearity. We measured very broad angular Schmidt spectra with the corresponding Schmidt numbers up to 229, which to the best of our knowledge is the highest ever reported angular Schmidt number.

\begin{figure}[t!]
\includegraphics{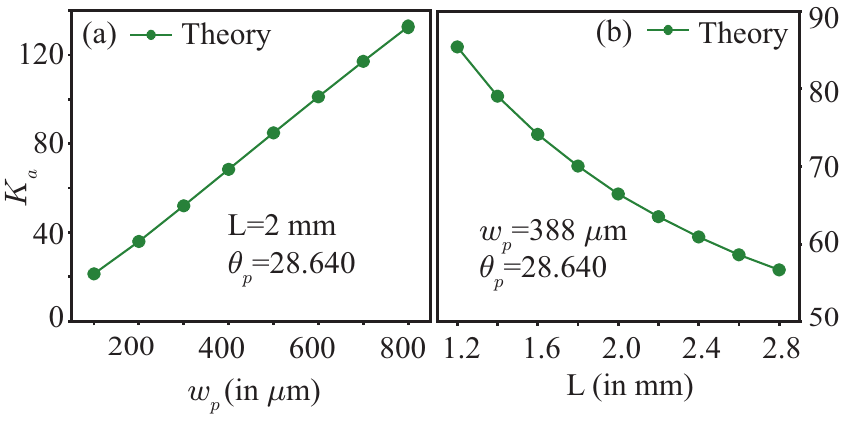}
\caption{(color online)  (a) and (b) Theoretical dependence of the angular Schmidt number on the width of the pump beam waist $w_p$ and crystal thickness $L$, respectively.}
\label{fig5}
\end{figure}

We find excellent agreement between the theory and experiment, except for extremely non-collinear conditions, in which case the experimentally measured Schmidt numbers are slightly lower than the theoretical predictions. The main reason for this discrepancy is the limited resolution of our EMCCD camera. In order to generate the azimuthal intensity plots, we use narrow angular region of interest (ROI) \cite{kulkarni2017natcomm}, the minimum possible size of which is fixed by the pixel size of the EMCCD camera. In the case of non-collinear down-conversion, the intensities in the interferograms are concentrated at regions away from the center. Therefore, the corresponding $\Delta \bar{I}_{\rm out}(\phi)$ plots have lesser angular resolution and thus they get estimated to be wider than their true widths. This results in a progressively lower estimate of the Schmidt numbers with increasing non-collinearity. Finally, we use Eq.~(\ref{angular_schmidt_spectrum2}) for studying how $w_p$ and $L$ affect the angular Schmidt number $K_a$. Figure \ref{fig5}(a) shows the theoretical dependence of $K_a$ on $w_p$ for fixed $L$, $\theta_p$ and $d$. Figure \ref{fig5}(b) shows the theoretical dependence of $K_a$ on $L$ for fixed $w_p$, $\theta_p$, and $d$. We find that $K_a$ increases as a function of $w_p$ while it decreases as a function of $L$. 

\section{Conclusion}

In summary, we have derived in this article an exact formula for the angular Schmidt spectrum of OAM-entangled photons produced by PDC. We have shown that our formula yields the true theoretical spectrum without any convergence issue as has been the case with the previously derived formulas. Furthermore, we have used our theoretical formulation to experimentally characterize the angular Schmidt spectrum for non-collinear phase matching in PDC. The results reported in this article can be very relevant for the ongoing intensive research efforts towards harnessing high-dimensional OAM entanglement for quantum information applications \cite{pors2011jo, erhard2018lsa}.

\section*{Acknowledgment}

We acknowledge financial support through grant no. IITK /PHY /20130008 from Indian Institute of Technology (IIT) Kanpur, India and through the research grant no. EMR/2015/001931 from the Science and Engineering Research Board (SERB), DST, Government of India.

%
%

\end{document}